\documentclass[aps,prb,letterpaper,twocolumn,showpacs,floatfix,superscriptaddress,longbibliography,amsmath,amsfonts,amssymb,preprintnumbers,citeautoscript]{revtex4-2}
\usepackage{CJK}
\usepackage{times}
\usepackage{graphicx}
\usepackage{amsmath}
\usepackage{amssymb}
\usepackage[colorlinks,linkcolor=blue,citecolor=blue,urlcolor=blue,hyperindex,pdfstartview=FitH,plainpages=false]{hyperref}
\usepackage{float}
\usepackage{lineno}
\usepackage[utf8]{inputenc}

\newcommand{\Tc}{$T_\text{c}$}

\newcommand{\MgB}{MgB$_2$}

\begin{document}

\begin{CJK*}{GBK}{}

\title{Decoupled interband pairing in a bilayer iron-based superconductor evidenced by ultrahigh-resolution ARPES}
\author{Shichong Wang}
\affiliation{Key Laboratory of Artificial Structures and Quantum Control (Ministry of Education), School of Physics and Astronomy, Shanghai Jiao Tong University, Shanghai 200240, China}
\affiliation{Beijing National Laboratory for Condensed Matter Physics, Institute of Physics, Chinese Academy of Sciences, Beijing 100190, China}
\author{Yuanyuan Yang}
\affiliation{Key Laboratory of Artificial Structures and Quantum Control (Ministry of Education), School of Physics and Astronomy, Shanghai Jiao Tong University, Shanghai 200240, China}
\author{Yang Li}
\affiliation{Beijing National Laboratory for Condensed Matter Physics, Institute of Physics, Chinese Academy of Sciences, Beijing 100190, China}
\affiliation{School of Physical Sciences, University of Chinese Academy of Sciences, Beijing 100190, China}
\author{Wenshan Hong}
\affiliation{Beijing National Laboratory for Condensed Matter Physics, Institute of Physics, Chinese Academy of Sciences, Beijing 100190, China}
\author{Huaxun Li}
\affiliation{School of Physics, Zhejiang University, Hangzhou 310058, China}

\author{Shaofeng Duan}
\affiliation{Beijing National Laboratory for Condensed Matter Physics, Institute of Physics, Chinese Academy of Sciences, Beijing 100190, China}
\affiliation{Key Laboratory of Artificial Structures and Quantum Control (Ministry of Education), School of Physics and Astronomy, Shanghai Jiao Tong University, Shanghai 200240, China}
\author{Lingxiao Gu}
\author{Haoran Liu}
\author{Jiongyu Huang}
\author{Jianzhe Liu}
\affiliation{Key Laboratory of Artificial Structures and Quantum Control (Ministry of Education), School of Physics and Astronomy, Shanghai Jiao Tong University, Shanghai 200240, China}
\affiliation{Beijing National Laboratory for Condensed Matter Physics, Institute of Physics, Chinese Academy of Sciences, Beijing 100190, China}
\author{Dong Qian}
\affiliation{Key Laboratory of Artificial Structures and Quantum Control (Ministry of Education), School of Physics and Astronomy, Shanghai Jiao Tong University, Shanghai 200240, China}
\affiliation{Collaborative Innovation Center of Advanced Microstructures, Nanjing University, Nanjing 210093, China}
\affiliation{Tsung-Dao Lee Institute, Shanghai Jiao Tong University, Shanghai 200240, China}
\author{Guanghan Cao}
\affiliation{School of Physics, Zhejiang University, Hangzhou 310058, China}
\affiliation{Collaborative Innovation Center of Advanced Microstructures, Nanjing University, Nanjing 210093, China}
\author{Huiqian Luo}
\affiliation{Beijing National Laboratory for Condensed Matter Physics, Institute of Physics, Chinese Academy of Sciences, Beijing 100190, China}
\author{Wentao Zhang}
\email{wentaozhang@iphy.ac.cn}

\affiliation{Beijing National Laboratory for Condensed Matter Physics, Institute of Physics, Chinese Academy of Sciences, Beijing 100190, China}
\affiliation{Key Laboratory of Artificial Structures and Quantum Control (Ministry of Education), School of Physics and Astronomy, Shanghai Jiao Tong University, Shanghai 200240, China}
\date {\today}
\begin{abstract}

We present direct experimental evidence of a weakly coupled multiband superconducting state in the bilayer iron-based superconductor ACa$_2$Fe$_4$As$_4$F$_2$ (A = K, Cs) via ultrahigh-resolution angle-resolved photoemission spectroscopy (ARPES). Remarkably, the K-containing compound exhibits two distinct transition temperatures, corresponding to two separate sets of bilayer-split bands, as evidenced by temperature-dependent superconducting gap and spectral weight near the Fermi energy, while its Cs counterpart displays conventional single transition behavior. These experimental observations are well described by the weakly coupled two-band model of Eilenberger theory, which identifies suppressed interband pairing interactions between the bilayer-split bands as the key mechanism. By exploring quantum phenomena in the weak-coupling limit within a multiband system, our findings pave the way for engineering exotic superconductivity via band-selective pairing control.

\end{abstract}
\maketitle
\end{CJK*}

The concept of multiband superconductivity, initially proposed through theoretical extensions of BCS theory to multiple electronic bands \cite{Leggett1966,Suhl1959}, has undergone profound evolution driven by experimental discoveries. Early multiband systems like V$_3$Si \cite{Hardy1953}, CeCu$_2$Si$_2$ \cite{Steglich1979}, NbSe$_3$ \cite{PhysRevLett.39.161}, and Sr$_2$RuO$_4$ \cite{Maeno1994} provided intriguing research platforms but failed to fully establish the modern paradigm of multiband superconductivity. A pivotal advancement emerged with \MgB, where spectroscopic techniques unambiguously resolved two distinct superconducting gaps associated with $\sigma$- and $\pi$-bands \cite{Nagamatsu2001,Iavarone2002,Souma2003,Mou2015}.  This breakthrough propelled theoretical frameworks exploring interband coupling mechanisms, revealing that phonon-mediated interactions between bands could cooperatively enhance the critical temperature (\Tc) \cite{Bouquet2001,Liu2001,Choi2002,Zehetmayer2003,Nicol2005}. Later studies established interband scattering---mediated by phonons, spin fluctuations, or other bosonic modes---as a critical determinant of superconducting phenomenology. This mechanism governs gap symmetry, \Tc~enhancement, and anomalous behaviors in thermodynamic responses \cite{Bouquet2001,Sologubenko2002,Zehetmayer2004}.  A central question in multiband superconductivity involves reconciling the coexistence of strong interband coupling (as in MgB$_2$) and weakly coupled regimes exhibiting gap decoupling, for which no unambiguous cases have been confirmed. Although measurements of London penetration depth in V$_3$Si \cite{Cho2022} and 2H-NbSe$_2$ \cite{Alshemi2025} provide indirect evidence of weakly coupled gaps, direct spectroscopic validation through angle-resolved photoemission or quasiparticle interference remains undetected. Identifying materials with weak interband pairing is critical to distinguish universal multiband behavior from material-specific coupling effects.

Iron-based superconductors, discovered in 2008 \cite{Kamihara2008}, have emerged as a paradigmatic platform for multi-band superconductivity. The iron-based superconductors ACa$_2$Fe$_4$As$_4$F$_2$ (A = K, Rb, Cs; referred to as 12442)  have emerged as structural analogs to bilayer cuprates, attracting significant attention due to their unique layered architecture, which features alternating double FeAs superconducting layers separated by insulating Ca$_2$F$_2$ blocks \cite{Wang2016,Wang2017}.
The layered configuration induces bilayer splitting, resulting in two distinct sets of Fermi surfaces centered at the $\Gamma$ point \cite{Wu2020}. Despite extensive experimental investigations of the 12442 systems, some controversies remain in the experimental observations. A nodal gap has been suggested by muon-spin rotation ($\mu$SR) and point-contact Andreev-reflection spectroscopy studies \cite{Smidman2018,Kirschner2018,Torsello2022}, whereas transport, optical, scanning tunneling microscope (STM), and angle-resolved photoemission spectroscopy (ARPES) experiments have demonstrated that 12442 exhibits a nodeless gap \cite{Huang2019,Xu2019,Duan2021,Wu2020,Li2024a}. Additionally, optical experiments have provided indications of a pseudogap \cite{Hao2022,Zhang2022}. 

Here, we present spectroscopic evidence of weak pairing scattering in the bilayer iron-based superconductor ACa$_2$Fe$_4$As$_4$F$_2$ utilizing  ultrahigh-resoultion ARPES. Specifically, with improved resolution, we observe distinct superconducting gaps in K12442 with two sets of transition temperatures (\Tc~= 33.5 K, $T_\text{c}^*\approx$ 22 K) in the bilayer-split bands, indicating weak interband coupling. In contrast, the Cs-containing sample Cs12442 exhibits a single transition at the bulk \Tc~= 30 K, consistent with strong interband coupling.
The aforementioned observations can be effectively explained by the different coupling strengths in the two-band model of Eliashberg theory. Furthermore, our high-resolution data reveal definitive evidence of nodeless superconducting gaps and establish the absence of pseudogap behavior near the Brillouin zone center.

\begin{figure*}
\centering\includegraphics[width = 2\columnwidth] {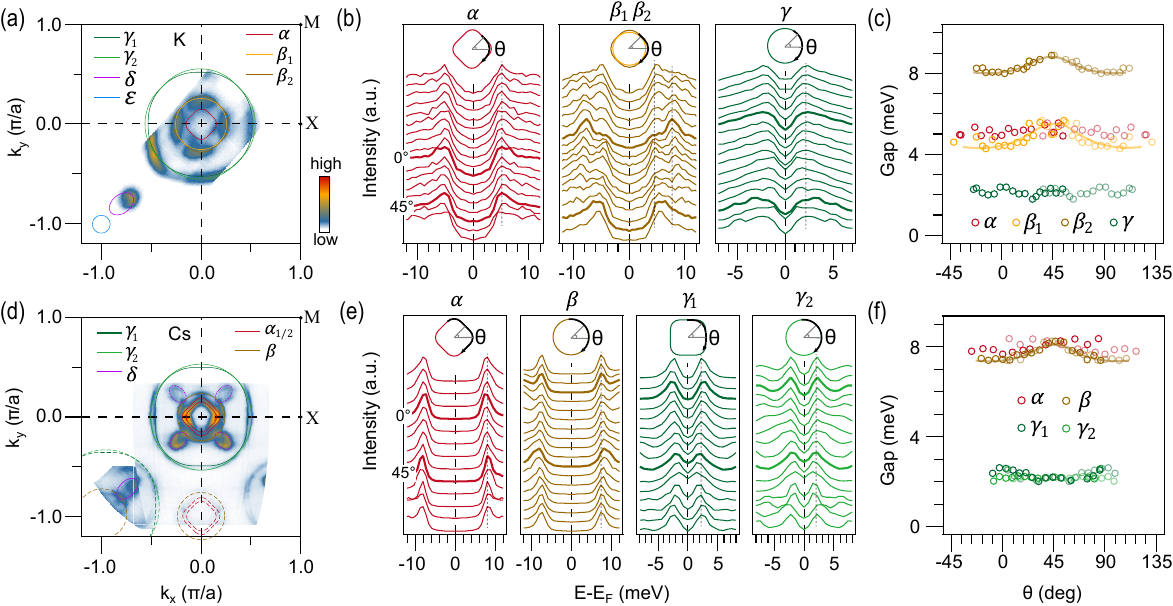}
\caption{
Fermi surface topology and superconducting gap characteristics of K12442 (upper panel) and Cs12442 (lower panel) measured at 4 K. (a) Photoemission contour for K12442 at a binding energy of 10 meV. (b) Momentum-dependent symmetrized EDCs corresponding to the hole bands at the $\Gamma$ point for K12442, with original EDCs shown in Supplemental Figs. S1 (b), (e) \cite{SupplMat}. (c)  Extracted energy gap size as a function of the marked angle shown in (b). The light hollow circles represent symmetrized data at $\theta$ = 45$^{\circ}$ to highlight the symmetry. (d)--(f) Corresponding measurements on Cs12442. Dashed lines of various colors indicate the corresponding folded bands.
}
\label{Fig1}
\end{figure*}

\begin{figure*}
\centering\includegraphics[width = 1.3\columnwidth] {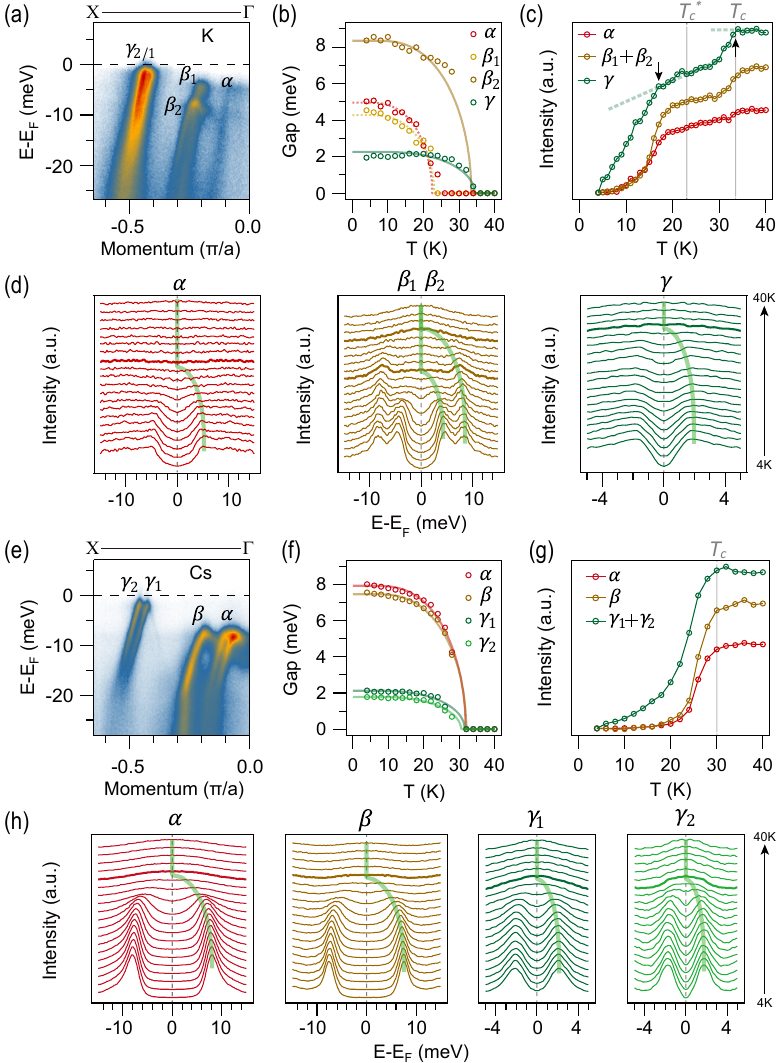}
\caption{
Temperature-dependent measurements along $\Gamma$-X in K12442 (a--d) and Cs12442 (e--h). (a), (e) High resolution photoemission spectra acquired near the Fermi energy at 4 K. (d), (h) Symmetrized EDCs for the hole bands near the $\Gamma$ point from 4 K to 40 K. Green guide lines highlight the temperature evolution of the superconducting gap. The determination of $k_F$ for the $\beta$ band is discussed in Supplemental Discussion \text{III}, and the original EDCs are shown in Supplemental Fig. S2 \cite{SupplMat}. (b), (f) Temperature dependence of the extracted energy gaps from panels (d) and (h), respectively. Solid and dashed curves represent fits to the BCS gap function. (c), (g) Temperature-dependent intensity integrated within $\pm$ 0.5 meV of the Fermi energy for individual bands.
} 
\label{Fig2}
\end{figure*}

Low-temperature ultrahigh-resolution ARPES measurements at 4 K confirm that bilayer splitting is a common characteristic in the 12442 system (see methods in Supplemental Material \cite{SupplMat}; see also Refs. \cite{Wang2020,Wang2016,Adroja2020,Hong2020,Huang2022,Yang2022,Duan2022} therein). In K12442, the Fermi surface consists of three hole-like pockets at the $\Gamma$ point: the $\alpha$ and $\beta$ bands (primarily from $d_{xz}$/$d_{yz}$ orbitals) and the $\gamma$ band (primarily from $d_{xy}$). A small electron-like $\varepsilon$ pocket is observed near M, surrounded by hole-like $\delta$ bands below the Fermi level [Fig. \ref{Fig1}(a)] \cite{Wu2020}. Pronounced splitting occurs along $\Gamma$-X, with minimal splitting along $\Gamma$-M [Fig. \ref{Fig2}(a) and Supplemental Fig. S1(c) \cite{SupplMat}]. Symmetrized energy distribution curves (EDCs) clearly resolve distinct energy gaps for the split $\beta$ bands: $\Delta_{\beta_1}$ $\sim$ 4.5 meV (bonding) and $\Delta_{\beta_2}$ $\sim$ 8.4 meV (antibonding), along with $\Delta_\alpha$ $\sim$ 5.0 meV and $\Delta_{\gamma}$ $\sim$ 2 meV [Figs. \ref{Fig1}(b) and (c)]. In Cs12442, band splitting is sharper [Fig. \ref{Fig1}(d)] \cite{Li2024}, with clearer surface reconstruction-induced folding effects (Supplemental Fig. S1(f) \cite{SupplMat}). The $\alpha$ and $\gamma$ bands exhibit more pronounced splitting than in the K-based compound, which is a consequence of stronger interlayer hopping and reduced inelastic scattering, with gap values of $\Delta_{\alpha1}$$\sim$ $\Delta_{\alpha2}$ $\sim$ 8.0 meV ($\Delta_{\alpha1}$: bonding band, $\Delta_{\alpha2}$: antibonding band), $\Delta_{\gamma_1}$$\sim$$\Delta_{\gamma_2}$ $\sim$ 2 meV, forming two sets of hole pockets [Fig. \ref{Fig1}(e)]. Despite these differences, both systems exhibit nearly isotropic superconducting gaps at the $\Gamma$ point across all bands except the $\beta$ band, which has a gap maximum along the zone diagonal [Figs. \ref{Fig1}(c) and (f)]. No nodal gap is detected in the hole pockets near $\Gamma$ in either K12442 or Cs12442.

Figure \ref{Fig2} reveals a stark contrast in the superconducting transitions of two samples. For K12442, data along the $\Gamma$-X direction show a dual-transition behavior. Symmetrized EDCs indicate the superconducting gap on the $\alpha$ and $\beta_1$ bands closes at a suppressed temperature ($T_\text{c}^*$$\sim$22 K), well below the bulk \Tc~of 33.5 K, while the $\beta_2$ and $\gamma$ bands close at \Tc~[Fig. \ref{Fig2}(d), Supplemental Fig. S3 and Discussion III for validation of the symmetrization \cite{SupplMat}]. This is confirmed by BCS gap fitting (Supplemental Discussion \text{IV} \cite{SupplMat}), which identifies two distinct transitions [Fig. \ref{Fig2}(b)]. The density of states (DOS) also shows two critical temperatures [Fig. \ref{Fig2}(c)], with the lower one slightly suppressed versus the gap-derived $T_\text{c}^*$, suggesting a subtle decoupling between spectral coherence and gap formation. A small discrepancy in the exact $T_\text{c}^*$ value between fitting methods indicates a residual gap persists above 22 K. This dual-transition phenomenon is robust across samples and thermal cycles (Supplemental Fig. S5) and is evidenced to be a bulk effect (Supplemental Discussion II; see also Refs. \cite{Duan2021, Shao2023, Liu2025} therein) \cite{SupplMat}. In sharp contrast, the Cs12442 sample exhibits a single, conventional BCS-like gap evolution at \Tc~for all bands [Figs. \ref{Fig2}(e)-(h)]. The absence of this dichotomy in Cs12442 highlights the unique role of suppressed interband coupling in K12442, likely tied to its distinct bilayer splitting.  No pseudogap signatures were observed near the $\Gamma$ point in either compound (Supplemental Discussion III and Fig. S4 \cite{SupplMat}).

\begin{figure}
\centering\includegraphics[width = 0.9\columnwidth] {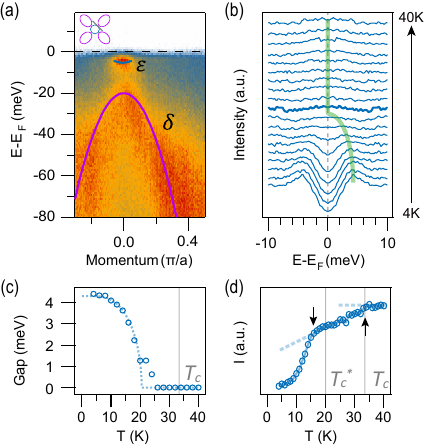}
\caption{
Measurements on K12442 near the M point. (a) Photoemission spectrum at 4 K for the momentum cut illustrated in the inset. The dispersions of the $\varepsilon$ and $\delta$ bands are schematized by the solid line. (b) Temperature-dependent symmetrized EDCs for the $\varepsilon$ band (bolded at $T_\text{c}^*$). (c) Temperature dependence of the superconducting gap in the $\varepsilon$ band fitted with a BCS function (dashed line).  (d) Spectral weight intensity of the $\varepsilon$ band within $\pm ~0.5$ meV of the Fermi level as a function of temperature.} 
\label{Fig3}
\end{figure}

In addition, we identified the electron-type $\varepsilon$ band crossing the Fermi level and the $\delta$ band situated approximately 20 meV below the Fermi level [Fig. \ref{Fig3}(a)]. At temperatures well below $T_\text{c}$, the $\varepsilon$ band exhibits a nearly flat dispersion, with a superconducting energy gap of around 4.4 meV [Figs. \ref{Fig3}(b) and (c)]. Consistent with the behavior observed in the $\alpha$ and $\beta_1$ bands, the temperature dependence of both the energy gap and spectral weight of the $\varepsilon$ band reveals a distinct transition at a lower temperature, $T_\text{c}^*$. Notably, the spectral weight intensity also shows a subtle transition at $T_\text{c}$, similar to the $\alpha$ band. The consistent observation of dual transitions near the M point provides additional evidence for weak interband coupling in this multiband superconductor.

We employ a simplified two-band weak-coupling model based on Eilenberger theory to interpret the experimental results \cite{Prozorov2011}. This approach has previously demonstrated consistency with experimental measurements of superfluid density and heat capacity in multiband superconductors such as MgB$_2$ and V$_3$Si \cite{Kogan2009,Prozorov2011}. In the two-band model, the energy gap equation $\Delta(T)$ can be numerically solved using the following equations (Supplemental Discussion \text{V} \cite{SupplMat}; see also \cite{Prozorov2011,Kogan2009,Bardeen1957,Eilenberger1968} therein):
\begin{small}
\begin{equation}
\begin{aligned}
\Delta_\nu &= \sum_{\mu=1,2} \lambda_{\nu\mu} \Delta_\mu [S + \ln (T/T_\text{c}) \\
&\quad - \sum_{m=0}^{\infty} ( \frac{1}{m + 1/2} - \frac{1}{\sqrt{( \frac{\Delta_\nu}{2\pi T} )^2 + (m + 1/2)^2}})].
\end{aligned}
\label{Eq1}
\end{equation}
\end{small}

Here, $\nu=1, 2$ represents the band index. The parameter $S$ is calculated using the coupling strength determined by $T_\text{c}$ (Supplemental Eqs. S10 and S12 \cite{SupplMat}). The DOS for the two bands, $n_1$ and $n_2$, satisfying $n_1 + n_2 = 1$, and the coupling strength $\lambda_{\nu\mu}$ is proportional to $n_\mu$. Typically, the interband coupling is significantly weaker than the intraband coupling ($\lambda_{12},\lambda_{21}\ll\lambda_{11},\lambda_{22}$). The temperature dependence of the energy gaps in the two-band system can be numerically solved within the framework of Eilenberger theory using Eq. \ref{Eq1}.

\begin{figure}
\centering\includegraphics[width = 0.8\columnwidth] {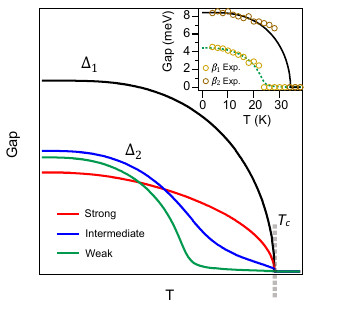}
\caption{
Simulated energy gaps as a function of temperature based on the Eilenberger two-band model for varying interband coupling strengths.
The interband coupling coefficients are defined as follows: weak coupling ($\lambda_{12} = \lambda_{21} = 0.001$), intermediate coupling ($\lambda_{12} = \lambda_{21} = 0.01$), and strong coupling ($\lambda_{12} = \lambda_{21} = 0.1$). The remaining parameters are fixed at $n_1 = n_2 = 0.5$, $\lambda_{11} = 1$, and $\lambda_{22} = 0.9$. A detailed discussion of the simulation methodology and results is provided in the Supplemental Discussion \text{V} \cite{SupplMat}. The inset shows the temperature-dependent gaps for the bilayer-split bands $\beta_1$ and $\beta_2$ in K12442 under weak interband coupling.}
\label{Fig4}
\end{figure}
 
Numerical solutions for the energy gap are shown in Fig. \ref{Fig4} for weak, intermediate, and strong coupling.
For the bilayer-split two bands, it is reasonable to assume $n_1=n_2$, giving $\lambda_{12}=\lambda_{21}$. The strong coupling scenario exhibits excellent agreement with experimental observations in Cs12442, consistent with prior ARPES studies on \MgB~\cite{Mou2015}. Conversely, the results for K12442 are consistent with weak coupling, indicating that replacing Cs with K reduces the interlayer coupling. A key observation is that even a minimal interband coupling strength ($\lambda_{12} = \lambda_{21} = 0.001$) is sufficient to suppress the emergence of a secondary transition temperature, unless the bands are entirely decoupled ($\lambda_{12} = \lambda_{21} = 0$). Consequently, in the weak coupling regime, the partial energy gap does not fully close prior to $T_\text{c}$, leaving a tiny residual gap between $T_\text{c}^*$ and $T_\text{c}$. This residual gap may be experimentally unresolvable, as suggested by the data in the inset of Fig. \ref{Fig4}. 

Lastly, we discuss the key factor that dominates the K-containing sample, so different from that of the Cs counterpart. We attribute the potential for weak coupling in K12442 to an anomalous reduction in the interaction potential during the pairing process. The interaction parameter $\lambda_{\nu\mu}$, which characterizes the coupling between bands, is proportional to the DOS ($n_{\mu}N(0)$) and the interaction potential ($V_{\nu\mu}$), defined as $\lambda_{\nu\mu} = n_{\mu} N(0)V_{\nu\mu}$. For the two samples in the 12442 system, the DOS $n_{\mu} N(0)$ is comparable, as the Fermi surface mappings in Figs. \ref{Fig1}(a) and \ref{Fig1}(d) give similar Fermi surface sheets. 
As discussed previously, the coupling strength $\lambda_{\nu\mu}$ is governed primarily by the interaction potential $V_{\nu\mu}$, which comprises the intraband electron pairing potential and the interband scattering potential between paired electrons \cite{Suhl1959}. 
Based on the above experimental observations, we conclude that the interband interaction potential $V_{\nu\mu}$ in the K sample should be significantly smaller than in the Cs sample. We note that the photoemission spectra for K12442 are broader, indicating heavier impurity scatterings than those in Cs12442. Our observations match a theoretical proposal where interband impurity scattering drives weak repulsive pairing and similar dual transitions in the temperature-dependent energy gap \cite{Liu2001,Stanev2014}. 
Therefore, we conclude that impurity scattering plays a key role in mediating the interband pairing scattering potential. Furthermore, the variation in interlayer spacing of the Fe$_2$As$_2$ layers, which is controlled by the size of the alkali atoms, also modulates this potential. Both of these factors are therefore crucial in determining the interband coupling strength $V_{\nu\mu}$.

In summary, using ultrahigh-resolution ARPES, we identify weakly coupled interband superconducting pairing in K12442, characterized by two distinct transition temperatures, while Cs12442 exhibits a single transition, consistent with strong coupling. These results are well described by a two-band Eilenberger theory with varying interband coupling strengths. This discovery provides a platform for studying the crossover between coupled and decoupled multiband superconductivity, reshaping the understanding of high-\Tc~ systems where multiband physics and correlations are intertwined. In addition, based on the experimental observation, we propose that reduced interband scattering protects against gap suppression, thereby enhancing superconductivity. Our studies open new avenues for exploring band-selective pairing and electronic correlations in unconventional superconductivity.

\begin{acknowledgements}
W. T. Z. acknowledges support from the National Key R\&D Program of China (Grants No. 2021YFA1401800 and No. 2021YFA1400202), the National Natural Science Foundation of China (Grants No. 12141404 and No. 12525404), and the Natural Science Foundation of Shanghai (Grants No. 22ZR1479700 and No. 23XD1422200). S. F. D. acknowledges support from the National Natural Science Foundation of China (Grants No. 12304178 and No. 12574166). D. Q. acknowledges support from the National Key R\&D Program of China (Grants No. 2022YFA1402400 and No. 2021YFA1400100) and the National Natural Science Foundation of China (Grant No. 12074248). H. Q. L. acknowledges support from the National Key R\&D Program of China (Grants No. 2023YFA1406100 and 2018YFA0704200) and the National Natural Science Foundation of China (Grants No. 11822411 and No. 12274444). G. H. C. acknowledges support from the National Key R\&D Program of China (Grants No. 2022YFA1403202 and 2023YFA1406101).
\end{acknowledgements}

\section*{Data Availability}
The data that support the findings of this article are not publicly available upon publication because it is not technically feasible, and/or the cost of preparing, depositing, and hosting the data would be prohibitive within the terms of this research project. The data are available from the authors upon reasonable request.

%

\clearpage
\onecolumngrid
\pagestyle{empty}
\begin{center}
\includegraphics[page=1, width=1.0\textwidth, height=1\textheight, keepaspectratio]{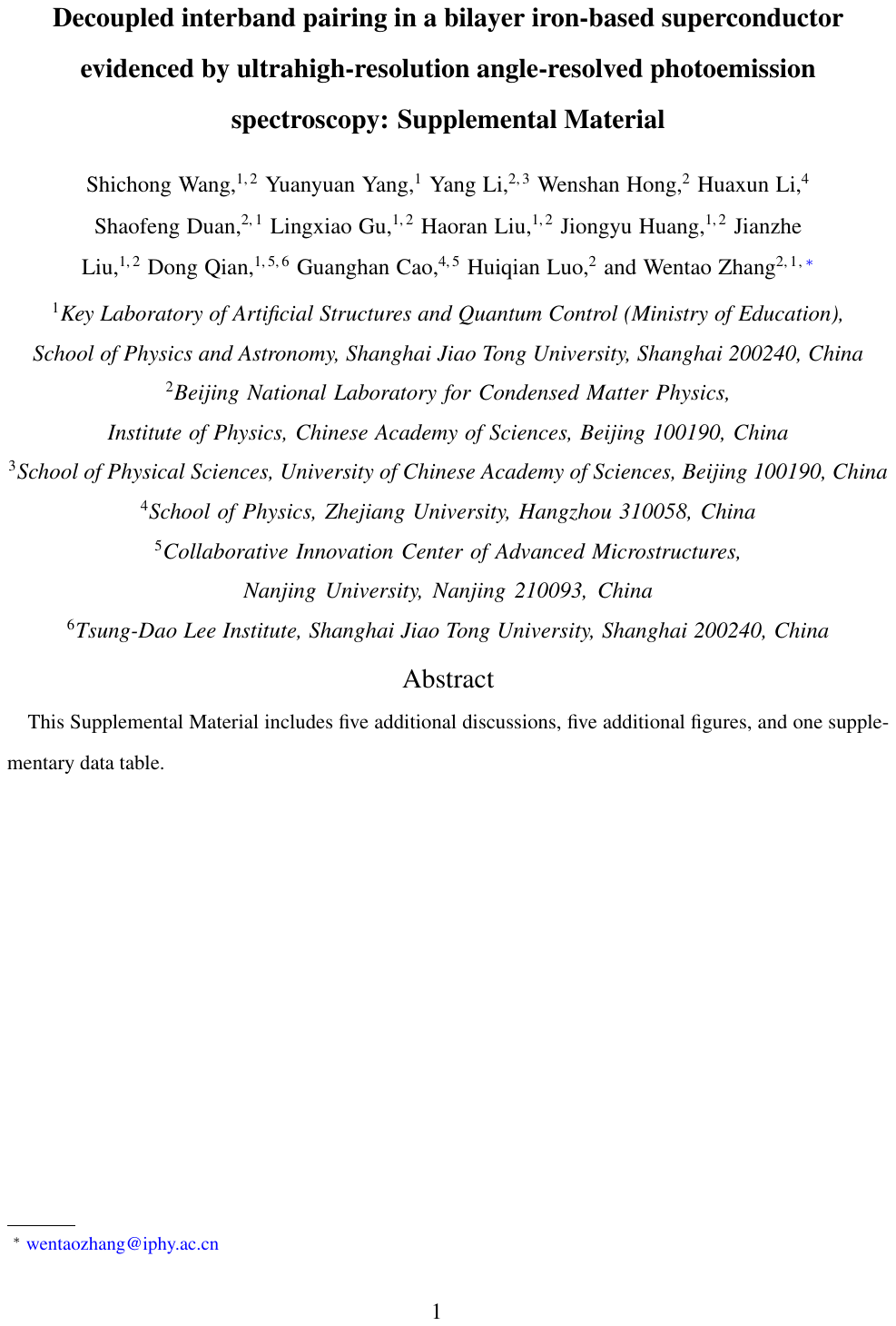}
\end{center}

\begin{center}
    \includegraphics[page=2, width=1.0\textwidth, height=1\textheight, keepaspectratio]{SM.pdf}
\end{center}
\clearpage

\begin{center}
    \includegraphics[page=3, width=1.0\textwidth, height=1\textheight, keepaspectratio]{SM.pdf}
\end{center}
\clearpage

\begin{center}
    \includegraphics[page=4, width=1.0\textwidth, height=1\textheight, keepaspectratio]{SM.pdf}
\end{center}
\clearpage

\begin{center}
    \includegraphics[page=5, width=1.0\textwidth, height=1\textheight, keepaspectratio]{SM.pdf}
\end{center}
\clearpage

\begin{center}
    \includegraphics[page=6, width=1.0\textwidth, height=1\textheight, keepaspectratio]{SM.pdf}
\end{center}
\clearpage

\begin{center}
    \includegraphics[page=7, width=1.0\textwidth, height=1\textheight, keepaspectratio]{SM.pdf}
\end{center}
\clearpage

\begin{center}
    \includegraphics[page=8, width=1.0\textwidth, height=1\textheight, keepaspectratio]{SM.pdf}
\end{center}
\clearpage

\begin{center}
    \includegraphics[page=9, width=1.0\textwidth, height=1\textheight, keepaspectratio]{SM.pdf}
\end{center}
\clearpage

\begin{center}
    \includegraphics[page=10, width=1.0\textwidth, height=1\textheight, keepaspectratio]{SM.pdf}
\end{center}
\clearpage

\begin{center}
    \includegraphics[page=11, width=1.0\textwidth, height=1\textheight, keepaspectratio]{SM.pdf}
\end{center}
\clearpage

\begin{center}
    \includegraphics[page=12, width=1.0\textwidth, height=1\textheight, keepaspectratio]{SM.pdf}
\end{center}
\clearpage

\begin{center}
    \includegraphics[page=13, width=1.0\textwidth, height=1\textheight, keepaspectratio]{SM.pdf}
\end{center}
\clearpage

\begin{center}
    \includegraphics[page=14, width=1.0\textwidth, height=1\textheight, keepaspectratio]{SM.pdf}
\end{center}
\clearpage

\end{document}